\RequirePackage{fix-cm}
\documentclass[twocolumn]{svjour3}          
\smartqed  
\usepackage{graphicx}
\begin{document}

\title{Antiadiabatic Phonons and Superconductivity in Eliashberg --
McMillan Theory}


\author{M.V. Sadovskii}


\institute{M.V. Sadovskii \at 
Institute for Electrophysics, Russian Academy of Sciences, 
Ural Branch,  Amundsen str. 106, Ekaterinburg 620016, Russia}

\date{Received: date / Accepted: date}

\maketitle

\begin{abstract}

The standard Eliashberg -- McMillan theory of superconductivity is essentially
based on the adiabatic approximation.  Here we present some
simple estimates of electron -- phonon interaction within Eliashberg -- McMillan
approach in non -- adiabatic and even antiadiabatic situation, when
characteristic phonon frequency $\Omega_0$ becomes large enough, i.e. comparable
or exceeding the Fermi energy $E_F$. We discuss the general definition of
Eliashberg -- McMillan (pairing) electron -- phonon  coupling constant $\lambda$,
taking into account the finite value of phonon frequencies. We show that the mass
renormalization of electrons is in general determined by different coupling
constant $\tilde\lambda$, which takes into account the finite width of conduction
band, and describes the smooth transition from the adiabatic regime to the region
of strong nonadiabaticity. In antiadiabatic limit, when $\Omega_0\gg E_F$, the new
small parameter of perturbation theory is
$\lambda\frac{E_F}{\Omega_0}\sim\lambda\frac{D}{\Omega_0}\ll 1$
($D$ is conduction band half -- width), and corrections to electronic spectrum
(mass renormalization) become irrelevant. However, the temperature of
superconducting transition $T_c$ in antiadiabatic limit is still determined by
Eliashberg -- McMillan coupling constant $\lambda$. We consider in detail the 
model with discrete set of (optical) phonon frequencies.
A general expression for superconducting transition temperature $T_c$ is
derived, which is valid in situation, when one (or several) of such phonons
becomes antiadiabatic. We also analyze the contribution of such phonons into
the Coulomb pseudopotential $\mu^{\star}$ and show, that antiadiabatic
phonons do not contribute to Tolmachev's logarithm and its value is
determined by partial contributions from adiabatic phonons only.

\end{abstract}

\section{Introduction}

Eliashberg -- McMillan superconductivity theory is currently the basis for
microscopic description of Cooper pairing and all general properties of conventional
superconductors \cite{Schr,Scal,Geb,Izy,All}.
It is is essentially based on adiabatic approximation and Migdal's theorem
\cite{Mig}, which allows to neglect the vertex corrections to
electron -- phonon coupling in typical metals. The actual small parameter of
perturbation theory is $\lambda\frac{\Omega_0}{E_F}\ll 1$, where
$\lambda$ is the dimensionless Eliashberg -- McMillan electron -- phonon
coupling constant, $\Omega_0$ is characteristic phonon frequency and $E_F$ is
Fermi energy of electrons. This leads to the widely accepted opinion,
that vertex corrections can be neglected even for the case of
$\lambda > 1$, due to the fact, that in common metal $\frac{\Omega_0}{E_F}\ll 1$ .
The possible breaking of Migdal's theorem for the case of
$\lambda\sim 1$ due to polaronic effects was widely discussed in the
literature \cite{Alx,Scl}. In the following we consider only the case of
$\lambda <1$ where we can safely neglect these effects \cite{Scl}.

Recently a number of superconductors was discovered, where the adiabatic
approximation is not necessarily valid, and characteristic frequencies of
phonons are of the order or even greater than Fermi energy. In this respect we
can mention single -- atomic layers of FeSe on the SrTiO$_3$ substrate
(FeSe/STO) \cite{UFN}, as well as record breaking hydride based
superconductors at high pressures \cite{Grk-Krs}. This is also the case in the
long -- standing puzzle of superconductivity in doped StTiO$_3$ \cite{Gork_0}.
The role of nonadiabatic phonons was recently analyzed in important papers by
Gor'kov \cite{Gork_1,Gork_2} within the standard BCS -- like weak -- coupling 
approach, and
directly addressed to these new superconductors. Here we review 
some further
estimates, derived by us in Refs. \cite{Sad_1,Sad_2} in the 
framework of
Eliashberg -- McMillan theory.

\section{Electron self -- energy and electron -- phonon coupling
constant}
\begin{figure}
\includegraphics[clip=true,width=0.5\textwidth]{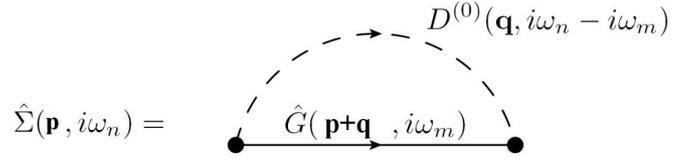}
\caption{Second -- order diagram for self -- energy. Dashed line --- phonon
Green's function $D^{(0)}$, continuous line --- electron Green's function
$G$ in Matsubara representation.}
\label{SE}
\end{figure}

Let us consider  first a metal in normal
(non superconducting) state, which is sufficient to introduce some basic
notions of Eliashberg -- McMillan theory \cite{Scal,Geb}.
The second -- order (in electron -- phonon coupling) diagram is shown in
Fig. \ref{SE}.  Making all calculations in finite temperature technique, after 
the analytic continuation from Matsubara to real frequencies 
$i\omega_n\to\varepsilon\pm i\delta$  and in
the limit of $T=0$ (i.e. $E_F\gg T$), the contribution of diagram in
Fig. \ref{SE} can be written \cite{Schr,Scal} as:
\begin{eqnarray}
\Sigma(\varepsilon,{\bf p})=\sum_{\bf p',\alpha}|g^{\alpha}_{\bf pp'}
|^2\Biggl\{\frac{f_{\bf p'}}{\varepsilon - \varepsilon_{\bf p'}
+\Omega^{\alpha}_{\bf p-p'}-i\delta}\nonumber\\
+ \frac{1-f_{\bf p'}}
{\varepsilon-\varepsilon_{\bf p'}-\Omega^{\alpha}_{\bf p-p'} + i\delta}\Biggr\}
\label{self-energy}
\end{eqnarray}
where in notations of Fig, \ref{SE} ${\bf p'=p+q}$.
Here $g^{\alpha}_{\bf p,p'}$ is Fr\"ohlich electron -- phonon coupling constant,
$\varepsilon_{\bf p}$ is electronic spectrum with energy zero taken at the Fermi
level, $\Omega^{\alpha}_{\bf q}$ represents the phonon spectrum, and 
$f_{\bf p}$ is Fermi distribution (step -- function at $T=0$).
In these expressions index $\alpha$ enumerates the branches of phonon spectrum,
which below is just dropped for brevity.

Now we can essentially follow the analysis, presented in Ref. \cite{Scal,Geb}.
Eq. (\ref{self-energy}) can be identically rewritten as:
\begin{eqnarray}
\Sigma(\varepsilon,{\bf p})=\int d\omega\sum_{\bf p'}|g_{\bf pp'}|^2
\delta(\omega-\Omega_{\bf p-p'})\times\nonumber\\
\times\Biggl\{\frac{f_{\bf p'}}
{\varepsilon - \varepsilon_{\bf p'}+\omega-i\delta}
+ \frac{1-f_{\bf p'}}
{\varepsilon - \varepsilon_{\bf p'}-\omega + i\delta}\Biggr\}
\label{self-energy_1}
\end{eqnarray}
To simplify calulations we can get rid of explicit momentum
dependencies here by averaging the matrix element of electron -- phonon
interaction over surfaces of constant energies, corresponding to
initial and final momenta ${\bf p}$ and ${\bf p'}$, which usually reduces to
the averaging over corresponding Fermi surfaces, as phonon scattering takes
place only within the narrow energy interval close to the Fermi level, with
effective width of the order of double characteristic frequency of phonons
$2\Omega_0$, and taking into account that in typical metals we always have
$\Omega_0\ll E_F$.

\begin{figure}
\includegraphics[clip=true,width=0.3\textwidth]{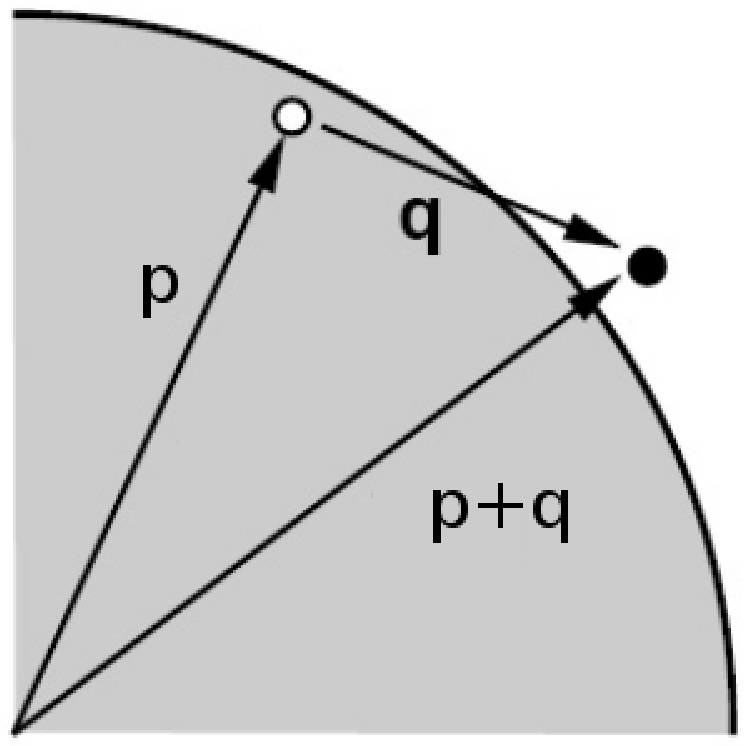}
\includegraphics[clip=true,width=0.4\textwidth]{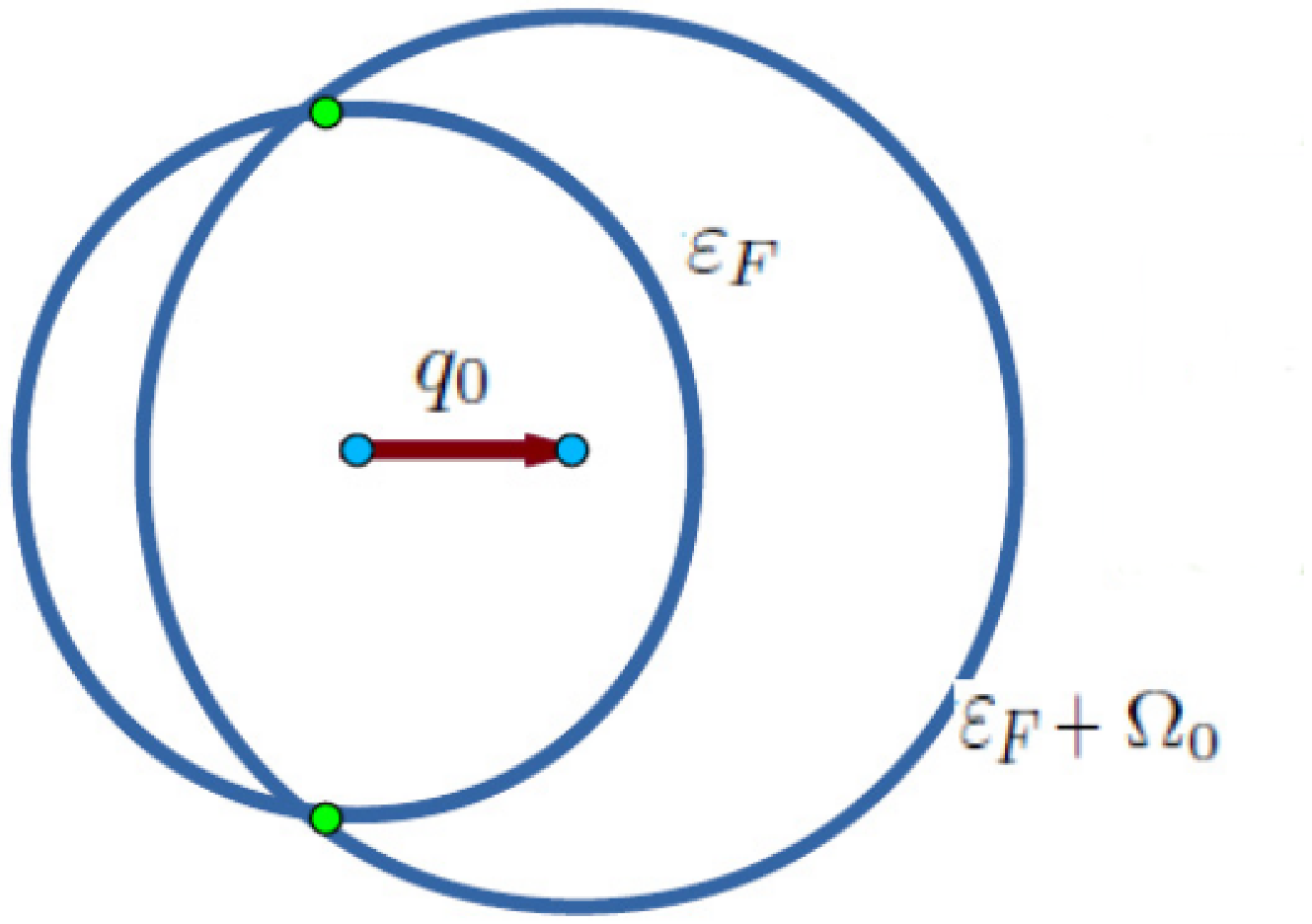}
\caption{(a) Elementary act of electron -- phonon  scattering in the vicinity
of the Fermi surface.
(b) Surfaces of constant energy for initial and final states of an electron
scattered by an optical phonon with energy comparable  to Fermi energy.
Non -- trivial contribution to the average of the matrix element in
(\ref{Elias_lambda}) or (\ref{e-ph-const}) comes here from the intersection
of these surfaces.}
\label{FS}
\end{figure}

This averaging can be achieved by the
following replacement in Eq. (\ref{self-energy_1}):
\begin{eqnarray}
|g_{\bf pp'}|^2\delta(\omega-\Omega_{\bf p-p'})\Longrightarrow\nonumber\\
\frac{1}{N(0)}
\sum_{\bf p}\frac{1}{N(0)}\sum_{\bf p'}
|g_{\bf pp'}|^2\delta(\omega-\Omega_{\bf p-p'})\delta(\varepsilon_{\bf p})
\delta(\varepsilon_{\bf p'})\nonumber\\
\equiv\frac{1}{N(0)}\alpha^2(\omega)F(\omega)
\label{Elias}
\end{eqnarray}
where in the last expression we have introduced the {\em definition} of
Eliashberg function $\alpha^2(\omega)$ and
$F(\omega)=\sum_{\bf q}\delta(\omega-\Omega_{\bf q})$ is the phonon density
of states.
In non -- adiabatic case, when phonon energy becomes comparable with or even
exceeds the Fermi energy, electron scattering is effective not only in the narrow
energy layer around the Fermi surface, but in much wider energy interval of the
order of $\Omega_0\sim E_F$. Then, for the case of initial $|{\bf p}|\sim p_F$
the averaging over ${\bf p'}$ in expression like (\ref{Elias}) should be done
over the surface of constant energy, corresponding to $E_F+\Omega_{\bf p-p'}$,
as is shown in Fig. (\ref{FS}). Now the Eq. (\ref{Elias}) is directly
generalized as:
\begin{eqnarray}
\noindent
|g_{\bf pp'}|^2\delta(\omega-\Omega_{\bf p-p'})\Longrightarrow\nonumber\\
\frac{1}{N(0)}
\sum_{\bf p}\frac{1}{N(0)}\sum_{\bf p'}
|g_{\bf pp'}|^2\times\nonumber\\
\times\delta(\omega-\Omega_{\bf p-p'})\delta(\varepsilon_{\bf p})
\delta(\varepsilon_{\bf p'}-\Omega_{\bf p-p'})\nonumber\\
\equiv\frac{1}{N(0)}\alpha^2(\omega)F(\omega)\nonumber\\
\label{Elias_1}
\end{eqnarray}
After the replacement like (\ref{Elias}) or (\ref{Elias_1}) the explicit
momentum dependence of the self -- energy disappears and in fact in the
following we are dealing with Fermi surface average of self -- energy
$\Sigma(\varepsilon)\equiv\frac{1}{N(0)}\sum_{\bf p}
\delta(\varepsilon_{\bf p})\Sigma({\varepsilon,{\bf p}})$, which is now written
as:
\begin{eqnarray}
\Sigma(\varepsilon)=\int d\varepsilon'\int d\omega\alpha^2(\omega)F(\omega)
\Biggl\{\frac{f(\varepsilon')}
{\varepsilon - \varepsilon'+\omega-i\delta} \nonumber\\
+ \frac{1-f(\varepsilon')}
{\varepsilon - \varepsilon'-\omega + i\delta}\Biggr\}
\label{self-energy_2}
\end{eqnarray}
This expression forms the basis of Eliashberg -- McMillan theory and determines
the structure of Eliashberg equations for the description of superconductivity.
Now the self -- energy is dependent only on frequency (and not on momentum)
and we can use the following simple expressions, relating mass renormalization
of an electron to the residue a the pole of the Green's function \cite{Diagr}:
\begin{equation}
Z^{-1}=1-\left.\frac{\partial\Sigma(\varepsilon)}{\partial\varepsilon}
\right|_{\varepsilon=0}
\label{Z_res}
\end{equation}
\begin{equation}
m^{\star}= \frac{m}{Z}=m\left.\left(1-\frac{\partial\Sigma(\varepsilon)}
{\partial\varepsilon}
\right|_{\varepsilon=0}\right)
\label{m_eff}
\end{equation}
Then from Eq. (\ref{self-energy_2}) by direct calculations we obtain:
\begin{equation}
-\left.\frac{\partial\Sigma(\varepsilon)}{\partial\varepsilon}
\right|_{\varepsilon=0}=
2\int_{0}^{\infty}\frac{d\omega}{\omega}
\alpha^2(\omega)F(\omega)\nonumber\\
\label{derivat_0}
\end{equation}
and introducing the dimensionless Eliashberg -- McMillan electron -- phonon
coupling constant as:
\begin{equation}
\lambda=2\int_{0}^{\infty}\frac{d\omega}{\omega}\alpha^2(\omega)F(\omega)
\label{lambda_Elias_Mc}
\end{equation}
we immediately obtain the standard expression for electron mass renormalization
due to electron -- phonon interaction:
\begin{equation}
m^{\star}=m(1+\lambda)
\label{mass_ren}
\end{equation}
The function $\alpha^2(\omega)F(\omega)$ in the expression for Eliashberg --
McMillan electron -- phonon coupling constant (\ref{lambda_Elias_Mc}) should be
calculated according to (\ref{Elias}) or (\ref{Elias_1}) depending on the
relation between Fermi energy $E_F$ and characteristic phonon frequency $\Omega_0$
As long as $\Omega_0\ll E_F$ we can use the
standard expression (\ref{Elias}), while in case of $\Omega_0\sim E_F$ we should
use (\ref{Elias_1}).

Using Eq. (\ref{Elias_1}) we can rewrite (\ref{lambda_Elias_Mc}) in the
following form:
\begin{eqnarray}
\lambda=\frac{2}{N(0)}\int\frac{d\omega}{\omega}
\sum_{\bf p}\sum_{\bf p'}
|g_{\bf pp'}|^2\times\nonumber\\
\times\delta(\omega-\Omega_{\bf p-p'})\delta(\varepsilon_{\bf p})
\delta(\varepsilon_{\bf p'}-\Omega_{\bf p-p'})
\label{Elias_lambda}
\end{eqnarray}
which gives the most general expression to calculate the electron -- phonon
constant $\lambda$, determining pairing in Eliashberg -- McMillan
theory. Implicitly this result was contained already in Ref. \cite{Allen}.
Below we shall present some simple estimates, based on this general
relation.

\section{Estimates of electron -- phonon coupling with
non -- adiabatic phonons}

Let us consider the simplest possible model of electrons interacting with a
single optical (Einstein -- like) phonon mode with high -- enough frequency
$\Omega_0$. The general qualitative picture of such
scattering is shown in Fig. \ref{FS}. In this case in Eq. (\ref{Elias_lambda})
the density of phonon states is simply $F(\omega)=\delta(\omega-\Omega_0)$.
Just for orientation we may take the possible momentum dependence of 
interaction with this optical phonon in the form proposed in Refs.
\cite{FeSe_ARPES_Nature,Dolg} to describe nearly ``forward'' scattering by 
optical phonons at FeSe/STO interface, as a possible mechanism of strong $T_c$
enhancement in this system:
\begin{equation}
g({\bf q})=g_{0}\exp(-|{\bf q}|/q_0),
\label{g-forw}
\end{equation}
where the typical value of $q_0\ll p_F$ ($p_F$ is the Fermi momentum) to ensure
the nearly ``forward'' nature of scattering. This model allows explicit estimates, 
which may illustrate the general situation.

Now we can write the dimensionless pairing constant of electron -- phonon
interaction in Eliashberg theory as:
\begin{equation}
\lambda=\frac{2}{N(0)\Omega_0}
\sum_{\bf p}\sum_{\bf q}
|g_{\bf q}|^2\delta(\varepsilon_{\bf p})
\delta(\varepsilon_{\bf p+q}-\Omega_0)
\label{e-ph-const}
\end{equation}
As in FeSe/STO with rather shallow conduction band
\cite{FeSe_ARPES_Nature,NPS_1,NPS_2}, where in fact we have $\Omega_0> E_F$,
the finite value of $\Omega_0$ in the second $\delta$-function here should be 
definitely taken into account.

We can make our estimates assuming the simplest linearized form of electronic
spectrum near the Fermi surface
($ v_F $ is Fermi velocity): $\varepsilon_{\bf p}\approx v_F (| {\bf p} | -p_F)$,
which allows us to perform all calculations analytically. Using
(\ref{g-forw}) in (\ref{e-ph-const}) and considering the two -- dimensional case,
after the calculation of all integrals we obtain \cite{NPS_2}:
\begin{equation}
\lambda=\frac{g_0^2a^2}{\pi^2v_F^2}K_1\left(\frac{2\Omega_0}{v_Fq_0}\right),
\label{lambda}
\end{equation}
where $K_1(x)$ is Bessel function of imaginary argument (McDonald function).
Using the asymptotic form of $K_1(x)$ and dropping a number of
irrelevant constants of the order of unity, we get:
\begin{equation}
\lambda\sim\lambda_0\frac{q_0}{4\pi p_F},
\label{lambda_0}
\end{equation}
for $\frac{\Omega_0}{v_Fq_0}\ll 1$, and
\begin{equation}
\lambda\sim\lambda_0\frac{\Omega_0}{\pi E_F}\sqrt\frac{v_Fq_0}{\Omega_0}
\exp\left(-\frac{2\Omega_0}{v_Fq_0}\right),
\label{lambda_1}
\end{equation}
for $\frac{\Omega_0}{v_Fq_0} \gg 1$.
Here we introduced the standard dimensionless electron -- phonon coupling
constant:
\begin{equation}
\lambda_0=\frac{2g_0^2}{\Omega_0}N(0),
\label{lamb_0}
\end{equation}
where $N(0)$ is the density of electronic states at the Fermi level per single
spin projection.

The result (\ref{lambda_0}) is by itself rather unfavorable for significant
$T_c$ enhancement in model under discussion, where $q_0\ll p_F$.
Even worse is the situation if we take into account the large values of
$\Omega_0$, as pairing constant becomes exponentially suppressed for
$\frac{\Omega_0}{v_Fq_0}> 1$, which is typical for FeSe/STO interface, where
$\Omega_0> E_F \gg v_Fq_0 $ \cite{UFN}. This makes the enhancement of $T_c$
due to interaction of FeSe electrons with optical phonons of STO rather
improbable, as was stressed in Ref. \cite{NPS_2}.

However, this is not our main point here. Actually, using (\ref{g-forw}) we can
also make estimates for generally more typical case, when the optical phonon 
scatters electrons not only in nearly ``forward'' direction, but in a wider 
interval of transferred momenta. To do that we have simply to use in 
Eq. (\ref{g-forw})the larger values of parameter $q_0$. Choosing e.g. 
$q_0\sim 4\pi p_F$ and using the low frequency limit of (\ref{lambda_0}) we 
immediately obtain $\lambda\approx\lambda_0$,
i.e. the standard result. Similarly, parameter $q_0$ can be taken of the order 
of inverse lattice vector $2\pi/a$ (where $a$ is the lattice constant). 
Then for $q_0\sim 2\pi/a$ from (\ref{lambda_0}) we obtain:
\begin{equation}
\lambda\sim\lambda_0\frac{1}{2p_Fa}\sim\lambda_0
\label{lambda_st}
\end{equation}
for the typical case of $p_F\sim 1/2a$. In general there always remains 
the dependence on the value of Fermi momentum and cutoff parameter (cf. similar 
analysis in Ref. \cite{Diagr}). These particular estimates are valid for the 
adiabatic case.

In antiadiabatic limit of (\ref{lambda_1}), assuming $q_0\sim p_F$ we 
immediately obtain:
\begin{equation}
\lambda\sim\frac{\sqrt{2}}{\pi}\lambda_0\sqrt\frac{\Omega_0}{E_F}
\exp\left(-\frac{\Omega_0}{E_F}\right),
\label{lambda_hifr}
\end{equation}
which simply signifies the effective interaction cutoff for $\Omega_0>E_F$
in the antiadiabatic limit. This fact was already noted by Gor'kov in
Refs. \cite{Gork_1,Gork_2}, where it was stressed that in antiadiabatic limit
the cutoff in the Cooper channel is determined not by the average phonon
frequency, but by Fermi energy.

\section{Antiadiabatic limit and mass renormalization}

Our discussion up to now implicitly assumed the conduction band of an infinite
width. However, it is obvious that in case of large enough characteristic phonon
frequency it may become comparable with conduction band width, which in typical 
metal case is of the order of Fermi energy $E_F$. Now we will show that in the 
strongly nonadiabatic (antiadiabatic) limit, when $\Omega_0\gg E_F\sim D$ 
(here $D$ is the conduction band half-width), we are in fact dealing with the 
situation, when there appears a new small parameter of perturbation theory
$\lambda D/\Omega_0\sim\lambda E_F/\Omega_0$.

Consider the case of conduction band of the finite width $2D$ with
constant density of states (which formally corresponds to two --
dimensional case). The Fermi level as always is considered as an origin of
energy scale and for simplicity we assume the case of half -- filled band.
Then (\ref{self-energy_2}) reduces to:
\begin{eqnarray}
\noindent
\Sigma(\varepsilon)=\int_{-D}^{D} d\varepsilon'\int d\omega\alpha^2(\omega)
F(\omega)\Biggl\{\frac{f(\varepsilon')}
{\varepsilon - \varepsilon'+\omega-i\delta}\nonumber\\
+ \frac{1-f(\varepsilon')}
{\varepsilon - \varepsilon'-\omega + i\delta}\Biggr\}=\nonumber\\
\noindent
=\int d\omega\alpha^2(\omega)F(\omega)
\Biggl\{\ln\frac{\varepsilon+D+\omega-i\delta}{\varepsilon-D-\omega+i\delta}
\nonumber\\
\noindent
-\ln\frac{\varepsilon+\omega-i\delta}{\varepsilon-\omega+i\delta}\Biggr\}
\label{A1}
\nonumber\\
\end{eqnarray}
For the model of a single optical phonon 
$F(\omega)=\delta(\omega-\Omega_0)$ and we immediately obtain:
\begin{eqnarray}
\Sigma(\varepsilon)=\alpha^2(\Omega_0)F(\Omega_0)\Biggl\{
\ln\frac{\varepsilon+D+\Omega_0-i\delta}{\varepsilon-D-\Omega_0+i\delta}
\nonumber\\
\noindent
-\ln\frac{\varepsilon+\Omega_0-i\delta}{\varepsilon-\Omega_0+i\delta}
\biggr\}\nonumber\\
\label{A2}
\end{eqnarray}
Correspondingly, from (\ref{A1}) we get:
\begin{equation}
-\left.\frac{\partial\Sigma(\varepsilon)}{\partial\varepsilon}
\right|_{\varepsilon=0}
=2\int_{0}^{\infty}d\omega\alpha^2(\omega)F(\omega)\frac{D}{\omega(\omega+D)}
\nonumber\\
\label{A5}
\end{equation}
and we can define the {\em new} generalized coupling constant as:
\begin{equation}
\tilde\lambda=2\int_{0}^{\infty}\frac{d\omega}{\omega}\alpha^2(\omega)
F(\omega)\frac{D}{\omega+D}
\label{A6}
\end{equation}
which for $D\to\infty$ reduces to the usual Eliashberg -- McMillan constant
(\ref{lambda_Elias_Mc}), while for $D\to 0$ ($D\ll\Omega_0$) it gives the 
``antiadiabatic'' coupling constant:
\begin{equation}
\lambda_D=
2D\int \frac{d\omega}{\omega^2}\alpha^2(\omega)F(\omega)
\label{derivata_b}
\end{equation}
Eq. (\ref{A6}) describes the smooth transition between the limits of wide and
narrow conduction bands. Mass renormalization in general case is determined by
$\tilde\lambda$:
\begin{equation}
m^{\star}=m(1+\tilde\lambda)
\label{mass_renrm}
\end{equation}

For the model of a single optical phonon with frequency $\Omega_0$ we have:
\begin{equation}
\tilde\lambda=\frac{2}{\Omega_0}\alpha^2(\Omega_0)\frac{D}{\Omega_0+D}
=\lambda\frac{D}{\Omega_0+D}= \lambda_D\frac{\Omega_0}{\Omega_0+D}
\label{A7}
\end{equation}
where Eliashberg -- McMillan constant is:
\begin{equation}
\lambda=2\int_{0}^{\infty}\frac{d\omega}{\omega}\alpha^2(\omega)F(\omega)=
\alpha^2(\Omega_0)\frac{2}{\Omega_0}
\label{lambda_Elias_Mc_opt}
\end{equation}
and $\lambda_D$ reduces to:
\begin{equation}
\lambda_D=2\alpha^2(\Omega_0)\frac{D}{\Omega_0^2}=2\alpha^2(\Omega_0)
\frac{1}{\Omega_0}\frac{D}{\Omega_0}=\lambda\frac{D}{\Omega_0}
\label{lamb_D}
\end{equation}
where in the last expression we explicitly introduced the new small parameter
$D/\Omega_0\ll 1$, appearing in strong antiadiabatic limit. Correspondingly,
in this limit we always have:
\begin{equation}
\lambda_D=\lambda\frac{D}{\Omega_0}\sim\lambda\frac{E_F}{\Omega_0}\ll\lambda
\label{lamb_D_Mc}
\end{equation}
so that for reasonable values of $\lambda$ (even up to a strong coupling region
of $\lambda\sim 1$) ``antiadiabatic'' coupling constant remains small.
Obviously, all vertex corrections here are also small, as was shown rather
long ago by direct calculations in Ref. \cite{Ikeda}.
Thus we come to an unexpected conclusion --- in the limit of strong
nonadiabaticity the electron -- phonon coupling becomes weak and we obtain a
kind of ``anti -- Migdal'' theorem.

Physically, the weakness of electron -- phonon coupling in strong nonadiabatic
limit is more or less clear --- when ions move much faster than electrons,
these rapid oscillation are just averaged in time as electrons can not
follow the very rapidly changing configuration of ions.

\section{Eliashberg equations and the temperature of superconducting transition}

All analysis above was performed for the normal state of a metal. Now let us
turn to the superconducting phase.
The problem arises, to what extent the results
obtained can be generalized for the case of a metal in superconducting state?
In particular, what coupling constant ($\lambda$ or $\tilde\lambda$) 
determines the temperature of superconducting transition $T_c$ in 
antiadiabatic limit? Let us analyze this situation within appropriate 
generalization of Eliashberg equations.

Taking into account that in antiadiabatic approximation vertex corrections are
are again irrelevant and neglecting the direct Coulomb repulsion, Eliashberg 
equations can be derived in the usual way by calculating the diagram of 
Fig. \ref{SE}, where electronic Green's function in superconducting state is 
taken in Nambu's matrix representation. For real frequencies this Green's
function is written in the following standard form \cite{Geb,Izy}:
\begin{equation}
G(\varepsilon,{\bf p})=\frac{Z(\varepsilon)\varepsilon\tau_0+\varepsilon_{\bf p}
\tau_3+Z(\varepsilon)\Delta(\varepsilon)\tau_1}{Z^2(\varepsilon)\varepsilon^2-Z^2(\varepsilon)
\Delta^2(\varepsilon)-\varepsilon^2_{\bf p}}
\label{G_Nambu}
\end{equation}
which corresponds to the matrix of self -- energy:
\begin{equation}
\Sigma(\varepsilon)=[1-Z(\varepsilon)]\varepsilon\tau_0+Z(\varepsilon)
\Delta(\varepsilon)\tau_1
\label{SE_Nambu}
\end{equation}
where $\tau_i$ are standard Pauli matrices, while functions of mass
renormalization $Z(\varepsilon)$ and energy gap $\Delta(\varepsilon)$ are
determined from solution of integral Eliashberg equations \cite{Geb,Izy}.
For us now it is sufficient to consider only the linearized Eliashberg 
equations, determining superconducting transition temperature $T_c$, which for 
the case of real frequencies are written as \cite{Geb,Izy}:
\begin{eqnarray}
[1-Z(\varepsilon)]\varepsilon =\int_{0}^{D}d\varepsilon'\int_{0}^{\infty}d\omega
\alpha^2(\omega)F(\omega)f(-\varepsilon')\times\nonumber\\
\times\left(\frac{1}{\varepsilon'+\varepsilon+\omega+i\delta}-
\frac{1}{\varepsilon'-\varepsilon+\omega-i\delta}\right)
\label{lin_Z}
\end{eqnarray}
\begin{eqnarray}
Z(\varepsilon)\Delta(\varepsilon)=\int_{0}^{D}\frac{d\varepsilon'}{\varepsilon'}
th\frac{\varepsilon'}{2T_c}Re\Delta(\varepsilon')\times
\nonumber\\
\times\int_{0}^{\infty}d\omega
\alpha^2(\omega)F(\omega)
\times\nonumber\\
\times\left(\frac{1}{\varepsilon'+\varepsilon+\omega+i\delta}+
\frac{1}{\varepsilon'-\varepsilon+\omega-i\delta}\right)
\label{lin_D}
\end{eqnarray}
In difference with the standard approach \cite{Izy}, we have introduced the finite
integration limits, determined by the (half)bandwidth $D$.  To simplify the
analysis we again assume the half--filled band of degenerate electrons in two 
dimensions, so that $D=E_F\gg T_c$, with constant density of states. 

Situation is considerably simplified \cite{Sad_1,Sad_2}, if we consider these 
equations in the limit of $\varepsilon\to 0$ and look for the 
solutions\footnote{To avoid
confusion note, that according to standard notations of Eliashberg -- McMillan
theory the renormalization factor $Z$ as defined here is just the inverse of a
similar factor defined in Eq. (\ref{Z_res}) for the normal state} $Z(0)=Z$ and
$\Delta(0)=\Delta$. Then from (\ref{lin_Z}) we obtain:
\begin{equation}
[1-Z]\varepsilon=
-2\varepsilon\int_{0}^{\infty}d\omega\alpha^2(\omega)F(\omega)\frac{D}{\omega
(\omega+D)}
\label{Z_sc_eq}
\end{equation}
and we get the mass renormalization factor as:
\begin{equation}
Z=1+\tilde\lambda
\label{Z_sc}
\end{equation}
where constant $\tilde\lambda$ was defined above in Eq. (\ref{A6}), which
for $D\to\infty$ reduces to the usual Eliasberg -- McMillan constant
(\ref{lambda_Elias_Mc}), while for $D$ significantly smaller than characteristic
phonon frequencies it gives the ``antiadiabatic'' coupling constant
(\ref{derivata_b}). Mass renormalization is again determined by this 
generalized coupling constant $\tilde\lambda$ as in Eq. (\ref{mass_renrm}).
In particular, in the strong antiadiabatic limit this renormalization is
quite small and determined by the limiting expression $\lambda_D$ given by
Eq. (\ref{derivata_b}).

Situation is quite different in Eq. (\ref{lin_D}). In the limit of 
$\varepsilon\to 0$, using (\ref{Z_sc}) we immediately obtain from (\ref{lin_D}) 
the following equation for $T_c$:
\begin{equation}
1+\tilde\lambda=2\int_{0}^{\infty}d\omega\alpha^2(\omega)F(\omega)
\int_{0}^{D}\frac{d\varepsilon'}{\varepsilon'(\varepsilon'+\omega)}
th\frac{\varepsilon'}{2T_c}
\label{Tc}
\end{equation}
where $\lambda$ is the {\em standard} Eliashberg -- McMillan coupling constant 
as defined above in Eq. (\ref{lambda_Elias_Mc}). Thus, in general case, 
{\em different} coupling constants determine mass renormalization and $T_c$.

Let us consider rather general model with discrete set of dispersionless
phonon modes (Einstein phonons). In this case the phonon density of states
is written as:
\begin{equation}
F(\omega)=\sum_{i}\delta(\omega-\Omega_i)
\label{Fwi}
\end{equation}
where $\Omega_i$ are discrete frequencies modeling the optical branches of
the phonon spectrum. Then from Eqs. (\ref{lambda_Elias_Mc}) and (\ref{A6}) we get:
\begin{equation}
\lambda=2\sum_i\frac{\alpha^2(\Omega_i)}{\Omega_i}\equiv\sum_i\lambda_i
\label{lamb_i}
\end{equation}
\begin{equation}
\tilde\lambda=2\sum_i\frac{\alpha^2(\Omega_i)D}{\Omega_i(\Omega_i+D)}
=\sum_i\lambda_i\frac{D}{\Omega_i+D}
\equiv\sum_i\tilde\lambda_i
\label{lamb_tild}
\end{equation}
Correspondingly, in this case:
\begin{eqnarray}
\alpha^2(\omega)F(\omega)=\sum_i\alpha^2(\Omega_i)\delta(\omega-\Omega_i)\nonumber\\
=\sum_i\frac{\lambda_i}{2}\Omega_i\delta(\omega-\Omega_i)
\label{El-Mc-discr}
\end{eqnarray}
The standard Eliashberg equation (in adiabatic limit) for such model were
consistently solved in Ref. \cite{KM}. For our purposes it is sufficient to
analyze only Eq. (\ref{Tc}), which takes now the following form:
\begin{equation}
1+\tilde\lambda=2\sum_i\alpha^2(\Omega_i)
\int_{0}^{D}\frac{d\varepsilon'}{\varepsilon'(\varepsilon'+\Omega_i)}
th\frac{\varepsilon'}{2T_c}
\label{Tc_opt}
\end{equation}
This equation is easily solved to obtain:
\begin{equation}
T_c\sim
\prod_i\left(\frac{D}{1+\frac{D}{\Omega_i}}\right)^{\frac{\lambda_i}{\lambda}}
\exp\left(-\frac{1+\tilde\lambda}{\lambda}\right)
\label{Tc_opt_i}
\end{equation}
In the simple case of two optical phonons with frequencies $\Omega_1$ and
$\Omega_2$ we have:
\begin{equation}
T_c\sim
\left(\frac{D}{1+\frac{D}{\Omega_1}}\right)^{\frac{\lambda_1}{\lambda}}
\left(\frac{D}{1+\frac{D}{\Omega_2}}\right)^{\frac{\lambda_2}{\lambda}}
\exp\left(-\frac{1+\tilde\lambda}{\lambda}\right)
\label{Tc_opt_2}
\end{equation}
where $\tilde\lambda=\tilde\lambda_1+\tilde\lambda_2$ and
$\lambda=\lambda_1+\lambda_2$. For the case of $\Omega_1\ll D$ (adiabatic phonon),
and $\Omega_2\gg D$ (antiadiabatic phonon) Eq. (\ref{Tc_opt_2}) is immediately
reduced to:
\begin{equation}
T_c\sim (\Omega_1)^{\frac{\lambda_1}{\lambda}}
(D)^{\frac{\lambda_2}{\lambda}}
\exp\left(-\frac{1+\tilde\lambda}{\lambda}\right)
\label{Tc_opt_ad_ant}
\end{equation}
Now we can see, that in the preexponential factor the frequency of antiadiabatic
phonon is replaced by band half--width (Fermi energy), which plays a role of the
cutoff for logarithmic divergence in Cooper channel in antiadiabatic limit
\cite{Sad_1,Gork_1,Gork_2}.

Our general result (\ref{Tc_opt_i}) gives the general expression for $T_c$ for
the model with discrete set of optical phonons, valid both in adiabatic and
antiadiabatic regimes and interpolating between these limits in intermediate
region.
Actually Eq. (\ref{Tc_opt_i}) can be easily rewritten as:
\begin{equation}
T_c\sim\langle\Omega\rangle\exp\left(-\frac{1+\tilde\lambda}{\lambda}\right)
\label{Tc_log}
\end{equation}
where we have introduced the average logarithmic frequency 
$\langle\Omega\rangle$ as:
\begin{equation}
\ln \langle\Omega\rangle=\ln\prod_i
\left(\frac{D}{1+\frac{D}{\Omega_i}}\right)^{\frac{\lambda_i}{\lambda}}
=\sum_i\frac{\lambda_i}{\lambda}\ln \frac{D}{1+\frac{D}{\Omega_i}}
\label{Omega_ln}
\end{equation}
In the limit of continuous distribution of phonon frequencies this last
expression reduces to:
\begin{equation}
\ln\langle\Omega\rangle = \frac{2}{\lambda}\int\frac{d\omega}{\omega}
\alpha^2(\omega)F(\omega)\ln\frac{D}{1+\frac{D}{\omega}}
\label{Omega_log}
\end{equation}
where $\lambda$ is given by the usual expression (\ref{lambda_Elias_Mc}).
Eq. (\ref{Omega_log}) generalizes the standard definition of average logarithmic 
frequency of Eliasberg -- McMillan theory \cite{Izy} for the case of finite 
bandwidth.
Obviously, it reduces to the standard expression in adiabatic limit of
phonon frequencies much lower than $D$, and gives $\langle\Omega\rangle\sim D$ 
in extreme antiadiabatic limit, when all phonon frequencies are much larger 
than $D$.

\section{Coulomb pseudopotential}

Up to now we have neglected the direct Coulomb repulsion of electrons, which in
the standard approach \cite{Schr,Scal,Geb,Izy,All} is described by Coulomb pseudopotential
$\mu^{\star}$, which is effectively suppressed by large Tolmachev's logarithm.
As we noted in Ref. \cite{Sad_1} antiadiabatic phonons actually suppress Tolmachev's
logarithm, which can probably lead to rather strong suppression of the temperature
of superconducting transition. To clarify this situation we consider the
simplified version of integral equation for the gap (\ref{lin_D}), writing 
it in the standard form:
\begin{equation}
Z(\varepsilon)\Delta(\varepsilon)=
\int_{0}^{D}d\varepsilon'K(\varepsilon,\varepsilon')\frac{1}{\varepsilon'}
th\frac{\varepsilon'}{2T_c}\Delta(\varepsilon')
\label{Tcc}
\end{equation}
where the integral kernel is a combination of two step -- functions:
\begin{eqnarray}
K(\varepsilon,\varepsilon')=\lambda\theta(\langle\Omega\rangle-|\varepsilon|)
\theta(\langle\Omega\rangle-|\varepsilon'|)\nonumber\\
-\mu\theta(D-|\varepsilon|)\theta(D-|\varepsilon'|)
\label{K2step}
\end{eqnarray}
where $\mu$ is the dimensionless (repulsive) Coulomb potential, while the
parameter $\langle\Omega\rangle$, determining the energy width of attraction region due to
phonons is determined by preexponential factor (average logarithmic frequency) 
of Eqs. (\ref{Tc_opt_i}),(\ref{Tc_log}).
\begin{equation}
\langle\Omega\rangle=\prod_i\left(\frac{D}{1+\frac{D}{\Omega_i}}\right)^{\frac{\lambda_i}
{\lambda}}
\label{tildaD}
\end{equation}
It is important that we always have $\langle\Omega\rangle < D$.
Eq. (\ref{Tcc}) is now rewritten as:
\begin{eqnarray}
Z(\varepsilon)\Delta(\varepsilon)=(\lambda-\mu)\int_{0}^{\langle\Omega\rangle}
\frac{d\varepsilon'}{\varepsilon'}th\frac{\varepsilon'}{2T_c}\Delta(\varepsilon')
\nonumber\\
-\mu\int_{\langle\Omega\rangle}^{D}\frac{d\varepsilon'}{\varepsilon'}\Delta(\varepsilon')
\label{Tccc}
\end{eqnarray}
Writing the mass renormalization due to phonons as:
\begin{equation}
Z(\varepsilon)=\left\{\begin{array}{c}
1+\tilde\lambda\quad\mbox{for}\quad\varepsilon<\langle\Omega\rangle \\
1\quad\mbox{for}\quad\varepsilon>\langle\Omega\rangle
\end{array}
\right.
\label{Zmassph}
\end{equation}
we look for the solution of Eq. (\ref{Tcc}) for $\Delta(\varepsilon)$, as usual,
in the following form \cite{Scal,Izy,All}:
\begin{equation}
\Delta(\varepsilon)=\left\{\begin{array}{c}
\Delta_1\quad\mbox{for}\quad\varepsilon<\langle\Omega\rangle \\
\Delta_2\quad\mbox{for}\quad\varepsilon>\langle\Omega\rangle
\end{array}
\right.
\label{Delta2step}
\end{equation}
Then Eq. (\ref{Tccc}) is transformed into the system of two homogeneous linear
equations for constants $\Delta_1$ and $\Delta_2$:
\begin{eqnarray}
(1+\tilde\lambda)\Delta_1=(\lambda-\mu)\ln\frac{\langle\Omega\rangle}{T_c}\Delta_1
-\mu\ln\frac{D}{\langle\Omega\rangle}\Delta_2\nonumber\\
\Delta_2=-\mu\ln\frac{\langle\Omega\rangle}{T_c}\Delta_1-\mu\ln\frac{D}{\langle\Omega\rangle}\Delta_2
\label{D1D2}
\end{eqnarray}
The condition of the existence of nontrivial solution here is:
\begin{equation}
1+\tilde\lambda=\left(\lambda-\frac{\mu}{1+\mu\ln\frac{D}{\langle\Omega\rangle}}\right)
\ln\frac{\langle\Omega\rangle}{T_c}
\label{Detsys}
\end{equation}
Then the transition temperature is given by:
\begin{equation}
T_c=\langle\Omega\rangle\exp\left(-\frac{1+\tilde\lambda}{\lambda-\mu^{\star}}\right)
\label{Tcccc}
\end{equation}
where the Coulomb pseudopotential is determined as:
\begin{equation}
\mu^{\star}=\frac{\mu}{1+\mu\ln\frac{D}{\langle\Omega\rangle}}=
\frac{\mu}{1+\mu\ln\prod_i\left(1+\frac{D}{\Omega_i}\right)^{\frac{\lambda_i}{\lambda}}}
\label{mustar}
\end{equation}
Now the phonon frequencies enter Tolmachev's logarithm as the product of
partial contributions, with its values determined also by corresponding coupling
constants. Similar structure of Tolmachev's logarithm was first obtained
(in somehow different model) in Ref. \cite{KMK}, where the case of frequencies
going outside the limits of adiabatic approximation was not considered.
In this sense, Eq. (\ref{mustar}) has a wider region of applicability.
In particular, for the model of two optical phonons with frequencies
$\Omega_1\ll D$ (adiabatic phonon) and $\Omega_2\gg D$, from Eq. (\ref{mustar})
we get:
\begin{equation}
\mu^{\star}=\frac{\mu}{1+\mu\ln\left(\frac{D}{\Omega_1}
\right)^{\frac{\lambda_1}{\lambda}}}=
\frac{\mu}{1+\mu\frac{\lambda_1}{\lambda}\ln\frac{D}{\Omega_1}}
\label{mstar}
\end{equation}
We can see, that the contribution of antiadiabatic phonon drops out of
Tolmachev's logarithm, while the logarithm itself persists, with its value
determined by the ratio of the band halfwidth (Fermi energy) to the frequency
of adiabatic (low frequency) phonon. The general effect of suppression of
Coulomb repulsion also persists, though it becomes somehow weaker due
to the partial interaction of electrons with corresponding phonon.
This situation is conserved also in the general case --- the value of
Tolmachev's logarithm and corresponding Coulomb pseudopotential is
determined by contributions of adiabatic phonons, while antiadiabatic phonons
drop out. Thus, in general case, situation becomes more favorable for
superconductivity, as compared to the case of a single antiadiabatic phonon,
considered in Ref. \cite{Sad_1}.

\section{Conclusions}

In present paper we have considered the electron -- phonon coupling in
Eliashberg -- McMillan theory, taking into account antiadiabatic phonons with
high enough frequency (comparable or exceeding the Fermi energy $E_F$).
The value of mass renormalization, in general case, was shown to be
determined by the new coupling constant $\tilde\lambda$, while the value of
the pairing interaction is always determined by the standard coupling constant
$\lambda$ of Eliashberg -- McMillan theory, appropriately generalized by taking
into account the finite value of phonon frequency  \cite{Sad_1}.
Mass renormalization due to strongly antiadiabatic phonons is in general small and
determined by the coupling constant $\lambda_D\ll\lambda$. In this sense, in the
limit of strong antiadiabaticity, the coupling of such phonons with electrons
becomes weak and corresponding vertex correction again become irrelevant 
\cite{Ikeda,Sad_1}, creating a kind of ``anti -- Migdal'' situation. 
This fact allows us
to use Eliashberg -- McMillan approach in the limit of strong antiadiabaticity.
In the intermediate region all our expressions just produce a
smooth interpolation between adiabatic and antiadiabatic limits.

The cutoff of pairing interaction in Cooper channel in antiadiabatic limit
becomes effective at energies $\sim E_F\sim D$, as was previously noted in
Refs. \cite{Gork_1,Gork_2,Sad_1}), so that corresponding phonons do not
contribute to Tolmachev's logarithm in Coulomb pseudopotential.
However, the large enough values of this logarithm (and corresponding smallness
of $\mu^{\star}$) can be guaranteed due to contributions from adiabatic 
phonons \cite{Sad_2}.

Note that above we have used rather simplified analysis of Eliashberg
equations. However, in our opinion, more elaborate approach, e.g. along the
lines of Ref. \cite{KM}, will not lead to qualitative change of our results.
Some simple estimates for FeSe/STO system, based on these results, can be found
in Refs. \cite{Sad_1,Sad_2}.

\begin{acknowledgements}
It was a pleasure and honor to contribute this work to Ted Geballe's Festschrift,
as his work in the field of superconductivity has influenced all of us for decades.
This work was partially supported by RFBR grant No. 17-02-00015 and
the program of fundamental research No. 12 of the RAS Presidium ``Fundamental
problems of high -- temperature superconductivity''.
\end{acknowledgements}

\end{document}